\documentclass[preprint]{aastex}
\usepackage{emulateapj5}
\usepackage{epsf}

\slugcomment{Submitted to ApJL, 30 Nov 2001}
\shortauthors{KEETON}
\shorttitle{MASS CLUMPS IN B1422+231}

\newcommand\hMsun{h^{-1}\,M_\odot}

\newcommand\reffig[1]{Fig.~\ref{fig:#1}}

\makeatletter
\newenvironment{tablehere}
  {\def\@captype{table}}
  {}
\newenvironment{figurehere}
  {\def\@captype{figure}}
  {}
\makeatother

\begin{document}

\title{Discovering Mass Clumps in Distant Galaxies with Lensing: \\
  The Case of B1422+231}
\author{Charles R.\ Keeton\altaffilmark{1}}
\affil{
  Astronomy and Astrophysics Department, University of Chicago \\
  5640 S.\ Ellis Ave., Chicago, IL 60637 \\
}
\email{ckeeton@oddjob.uchicago.edu}
\altaffiltext{1}{Hubble Fellow}

\begin{abstract}
Substructure in distant gravitational lens galaxies can be detected
because it alters the brightnesses and shapes of the lensed images.
The optical and radio flux ratios in the four-image lens B1422+231
imply that there is a $\sim 10^{4}$--$10^{7}\ \hMsun$ mass clump in
the lens galaxy in front of image A; this is the first constraint on
the mass of a particular clump lying in a distant galaxy and detected
by its mass.  The data also indicate that a small clump, perhaps a
star, is passing in front of image B and making the optical flux
ratios variable.  Both of these hypotheses can be tested with new
observations.  B1422+231 demonstrates how data at different
wavelengths can be used in individual lenses to probe individual
mass clumps in distant galaxies.
\end{abstract}

\section{Introduction}

In hierarchical models of structure formation, the amount of
substructure in dark matter halos provides an important test of
the nature of the dark matter.  For cold dark matter (CDM), mass
clumps can survive the merger process and halos are predicted to
be lumpy (Klypin et al.\ 1999; Moore et al.\ 1999); while for
alternate types of dark matter (e.g., warm or self-interacting),
mass clumps are disrupted during mergers and halos are predicted
to be much more smooth (Spergel \& Steinhardt 2000; Colin,
Avila-Reese \& Valenzuela 2000; Bode, Ostriker \& Turok 2001).
In the Local Group, the number of dwarf galaxy satellites is
much smaller than the number of subhalos predicted by CDM, which
has been interpreted as a potentially profound problem with CDM
(Klypin et al.\ 1999; Moore et al.\ 1999).  This conclusion is
not unambiguous, however.  Astrophysical processes such as
photoionization can quench star formation in small halos
(Bullock, Kravtsov \& Weinberg 2000), so the number of detectable
satellite galaxies may under-represent the amount of substructure
in galaxy halos.

Gravitational lensing offers a better test for substructure because
it is directly sensitive to mass.  The image brightnesses in strongly
lensed systems are very sensitive to small-scale structure in the
lens galaxy.  Any mass clump, such as a globular cluster, gas cloud,
or satellite galaxy, can alter the brightness of an image by an
arbitrary amount (relative to a lens where the mass is smoothly
distributed; Mao \& Schneider 1998; Metcalf \& Madau 2001; Chiba
2001; Dalal \& Kochanek 2001).  Even a clump as small as a star
can perturb the images of small sources ($\lesssim 0.1$ pc; e.g.,
Chang \& Refsdal 1979); in this case, the star's orbital motion
can also induce variability in the image brightness with a time
scale of order months (e.g., Irwin et al.\ 1989).  I use the term
``sub-lensing'' for the generic phenomenon of strongly lensed images
being perturbed by substructure in the lens galaxy, and reserve the
term ``microlensing'' for events in which the time variability is
detectable.

The four-image lens B1422+231 is an example of a system where
sub-lensing appears to be important.  The flux ratios between the
images depend on both wavelength and time in a way that is
inconsistent with smooth lens models (Keeton, Kochanek \& Seljak
1997; Mao \& Schneider 1998; Metcalf \& Zhao 2001; also see \S 2).
I propose that the wavelength and time dependence imply two
distinct sub-lensing events occurring in this system: a relatively
massive subhalo in front of image A makes the radio flux ratios
differ from the optical flux ratios; and a small object, perhaps
a star, passing in front of image B makes the optical flux ratios
change with time.  The sub-lensing explanation for image A has
been considered before (Mao \& Schneider 1998; Chiba 2001; Dalal
\& Kochanek 2001), but only using a statistical treatment to
determine whether a plausible collection of mass clumps could
explain the radio flux.  Here I examine the hypothesis in more
detail, using joint fits to the radio and optical data to obtain
evidence for, and constraints on, particular objects acting as the
perturbers.  Section 2 reviews the data for B1422+231 from the
literature, Section 3 discusses modeling methods, Section 4 presents
results from the models, and Section 5 offers conclusions.

\section{Data}

B1422+231 consists of a radio-loud quasar at redshift $z_s=3.62$
that is lensed into four images by an elliptical galaxies at
redshift $z_l=0.34$ (Patnaik et al.\ 1992). The lens galaxies lies
in a poor group of galaxies that contributes an important tidal
shear to the lensing potential (Hogg \& Blandford 1994; Kundi\'c
et al.\ 1997). The shapes and polarizations of the images inferred
from VLBA observations are fully consistent with lensing (Patnaik
et al.\ 1999). For the lensed images, high-precision astrometry
from the VLBA maps yields relative positions with error ellipses
roughly 0.2 mas by 0.04 mas (Patnaik et al.\ 1999).  HST imaging
yields the relative position of the lens galaxy with an uncertainty
of 4 mas (Falco et al., in prep.).

Photometry of the four images has been obtained numerous times at
several radio frequencies and in various optical and near-IR
passbands. The most useful data are flux ratios between the images,
which are shown in \reffig{lcrv1}. Flux ratios are independent of
the intrinsic flux and variability of the source,\footnote{Provided
that the time scale for intrinsic variability is long compared with
the time delay between the images, which is true for B1422+231
(e.g., Patnaik \& Narasimha 2001).} so they should be independent
of both time and wavelength in simple lens models (see Schneider,
Ehlers \& Falco 1992). Contrary to expectations, there is a
significant difference between A/C(radio) and A/C(optical) that
persists over time.  The radio flux ratio has long presented
 problems for lens models (e.g., Keeton et al.\ 1997); rather
than being a failure of imagination among modelers, the problem
is generic to smooth lens models (Mao \& Schneider 1998). The flux
ratios A/C(optical), A/C(radio), and B/C(radio) are all consistent
(at better than 95\% confidence) with being constant over at least
a 6-year baseline,\footnote{Patnaik \& Narasimha (2001) claim to
see systematic variations in the 15 GHz radio fluxes during
March--September 1994, but the putative variations are smaller
than the errorbars.} while B/C(optical) appears to have declined
in recent years. The data therefore reveal two puzzles in B1422+231.
First, why is the A/C flux ratio different at radio and optical
wavelengths?  Second, why is the B/C flux ratio constant at radio
wavelengths and variable at optical wavelengths?  I propose that
both puzzles can be solved by invoking small-scale structure in
the lens galaxy, as explained in \S 4.

\begin{figurehere}
\centerline{\epsfxsize=3.2in \epsfbox{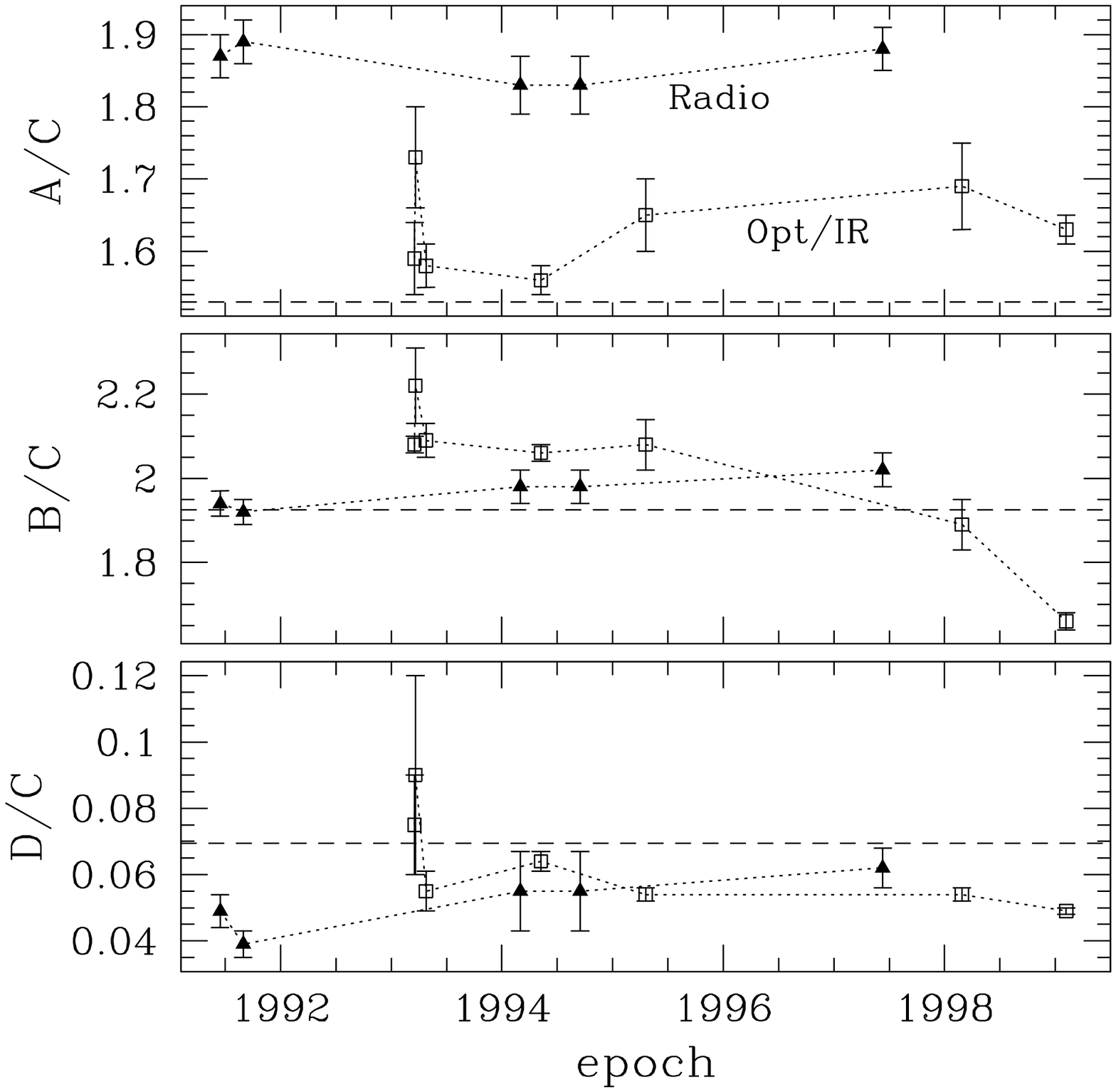}}
\caption{
Flux ratios for B1422+231 as a function of time. The dashed lines
indicate the flux ratios predicted by the macromodel for the lens
(see \S 3). The optical/near-IR data come from Remy et al.\ (1993),
Hammer et al.\ (1995), Yee \& Bechtold (1996), Impey et al.\
(1996), and Falco et al.\ (in prep.). The radio data come from
Patnaik et al.\ (1992, 1999) and Patnaik \& Narasimha (2001).
Differential extinction has little effect on the flux ratios
(Falco et al.\ 1999).
}\label{fig:lcrv1}
\end{figurehere}
\vspace{0.2cm}

\section{Modeling Techniques}

The lensing analysis begins with a smooth lens model, or
macromodel, fit to the global properties of the lens. I use a
fiducial model consisting of a singular isothermal ellipsoid for
the lens galaxy and an external shear to represent the tidal
perturbation from the group around the lens galaxy. The model is
constrained using the position data only (because the fluxes are
the subject of the sub-lensing analysis). The model gives a very
good fit; the image positions are fit arbitrarily well, and the
galaxy position is offset by
$(\Delta\alpha,\Delta\delta)=(-3.4,-5.2)$ mas giving a total
$\chi^2=2.4$ for two degrees of freedom. The best-fit lens galaxy
has an ellipticity $e=0.31\pm0.02$ at position angle
$\theta_e=-56\fdg1\pm0\fdg2$, with an external shear
$\gamma_{\rm ext}=0.164\pm0.005$ at position angle
$\theta_{\gamma,{\rm ext}}=-52\fdg60\pm0\fdg03$. (All uncertainties
are at 95\% confidence.) Even though the model was not constrained
by the image flux ratios, it agrees well with the B/C and D/C flux
ratios and slightly underpredicts the A/C optical flux ratio (see
\reffig{lcrv1}).  Table 1 gives the convergence $\kappa$,
the shear amplitude $\gamma$ and direction $\theta_\gamma$, and
the magnification $\mu$ predicted by the macromodel at each image.
Other macromodels can be found that predict different values for
the convergence and shear at each image.  Changes to the macromodel
would therefore modify the detailed quantitative results from the
sub-lensing analysis --- but would not change the main conclusion
that sub-lensing is at work in B1422+231.

In a sub-lensing analysis a mass clump is added to the macromodel
near one of the images, and the new lens equation is solved to find
the properties of the perturbed image. (Clumps lying far from the
image can be considered to be part of the macromodel and need not
be treated explicitly; see Metcalf \& Madau 2001.)  The new model
is evaluated with a $\chi^2$ statistic defined from the flux ratio
data in \reffig{lcrv1}.  I consider two types of mass clumps: a
point mass representing highly concentrated clumps, such as individual
stars or globular clusters; and a singular isothermal sphere (SIS)
representing more extended clumps, such as dwarf galaxy satellites.
Other clump models are possible, but these two are sufficient for
demonstrating that sub-lensing can explain the puzzles in B1422+231,
and for illustrating how the results depend on the clump model.

I assume that the radio and optical emission sources are coincident
but have different sizes. Combining the VLBA image shapes (Patnaik
et al.\ 1999) with the macromodel suggests that the radio source
is roughly circular with a FWHM of $\sim0.4$ mas. The continuum
optical emission region in quasars is thought to be $\sim 10^{15}$ cm
in size (e.g., Rees 1984; Wyithe et al.\ 2000), corresponding to
$\lesssim 10^{-4}$ mas for the source in B1422+231.

\vspace{2cm}
\begin{tablehere}
\begin{center}
{\sc Table 1. Macromodel Results} \\
\begin{tabular}{crrrr}
\tableline\tableline
\multicolumn{1}{c}{Image} &
\multicolumn{1}{c}{$\kappa$} &
\multicolumn{1}{c}{$\gamma$} &
\multicolumn{1}{c}{$\theta_\gamma$ ($^\circ$)} &
\multicolumn{1}{c}{$\mu$} \\
\tableline
 A & $0.384$ & $0.476$ & $-29.0$ & $ 6.57$ \\
 B & $0.471$ & $0.634$ & $-49.5$ & $-8.26$ \\
 C & $0.364$ & $0.414$ & $-83.1$ & $ 4.29$ \\
 D & $1.863$ & $2.025$ & $-55.1$ & $-0.30$ \\
\tableline
\label{tab:macro}
\end{tabular}
\end{center}
Note ---
The angles $\theta_\gamma$ are quoted as position angles measured
East of North.
\end{tablehere}
\vspace{0.2cm}

\begin{figurehere}
\centerline{\epsfxsize=3.2in \epsfbox{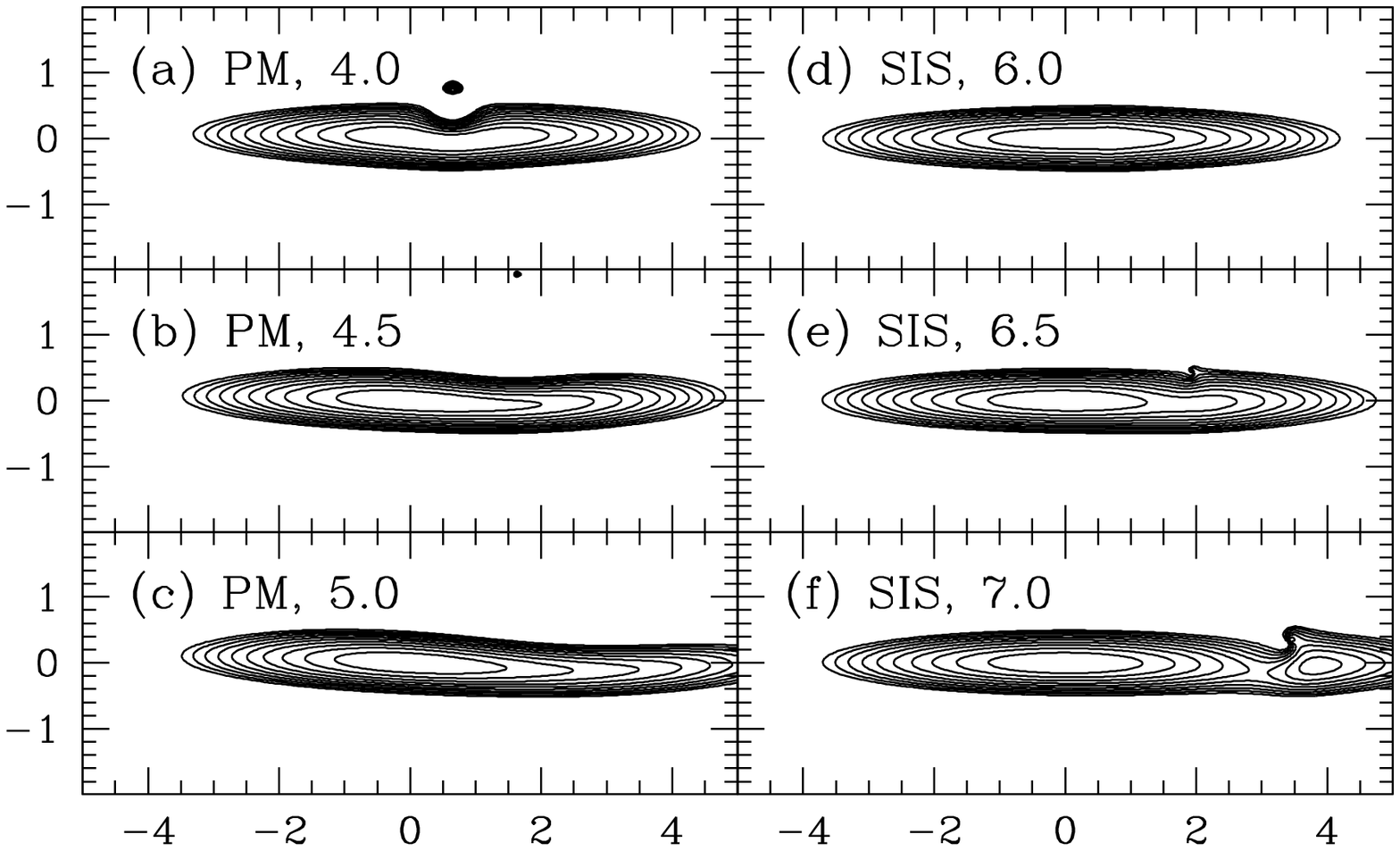}}
\caption{
Maps of the radio images predicted by various sub-lensing models
for image A, assuming infinite resolution.  Results are shown for
models with a point mass clump (left) and an SIS clump (right);
each panel gives the clump mass as $\log M$ (in $\hMsun$).  The
axes are labeled in mas, and the contours are spaced by 0.2 dex.
}\label{fig:maps}
\end{figurehere}
\vspace{0.2cm}

\section{Sub-Lensing Results}

\subsection{Image A}

In the sub-lensing picture for image A, a relatively large mass
clump perturbs the radio flux but not the optical flux. Two
conditions are required for the picture to work. First, the clump
must be massive enough to have an Einstein radius comparable in
size to the radio source. Second, the clump must be located at a
position where it can affect the radio image but not the optical
image --- which is possible because the radio image (size
$\sim 2.1 \times 0.4$ mas; Patnaik et al.\ 1999) is so much larger
than the optical image (size $\lesssim 10^{-4}$ mas).

Small clumps cannot produce strong enough perturbations to explain
the difference between the radio and optical flux ratios, so fits
to the data yield lower limits on the clump mass. For fiducial point
mass and SIS clumps, the mass limits are (at 95\% confidence)
\begin{equation}
  M > \cases{
0.9 \times 10^{4}\ (R_{src}/0.4\mbox{ mas})^2\ \hMsun & PM \cr
1.0 \times 10^{6}\ (R_{src}/0.4\mbox{ mas})^{3/2}\ \hMsun & SIS \cr
  }
\end{equation}
In this mass range, the time scale for variability due to motion
of the clump is several hundred years or longer.  The mass limits
depend on the clump model because, for a given clump mass, highly
concentrated point mass clumps produce larger changes in the
lensing potential than more extended SIS clumps.  Above these
lower limits, the flux data leave a degeneracy between the mass
and position of the clump. A more massive clump produces a
stronger perturbation, but it can still explain the fluxes if it
is moved to a lower surface brightness region of the radio image
(keeping the total radio flux fixed).

The mass degeneracy can be broken by adding information about the
shape of the radio image. \reffig{maps} shows that the image shape
depends on the clump mass; more massive clumps produce more
substantial distortions. Existing 1 mas resolution VLBA maps do not
show visible distortions of the type seen in Figures \ref{fig:maps}c
or \ref{fig:maps}f (Patnaik et al.\ 1999), so they imply that the
clump mass is not high,
\begin{equation}
  M \lesssim \cases {
    10^{5}\ (R_{src}/0.4\mbox{ mas})^2\ \hMsun & PM \cr
    10^{7}\ (R_{src}/0.4\mbox{ mas})^{3/2}\ \hMsun & SIS \cr
  }
\end{equation}
Higher resolution maps will do one of two things. If they do not
reveal any shape distortions in image A, they will disprove the
sub-lensing hypothesis as an explanation for the anomalous radio
flux ratio in B1422+231. However, if they do show distortions in
image A that are not seen in the other images, they will prove the
existence of a clump in front of image A. They will permit more
sophisticated models to use the image shape to determine the
position and mass of the clump. Moreover, because the image shape
depends on the nature of the clump (e.g., point mass or SIS; see
\reffig{maps}), it will allow models to determine the mass
distribution within the clump. This is an important general result:
flux data alone produce mass constraints that depend on the type of
clump assumed; but image shapes provide enough additional
constraints to determine clump properties like size and density
(see Metcalf \& Madau 2001 for more discussion).

\subsection{Image B}

For image B, sub-lensing must produce a variable optical flux ratio
without significantly affecting the radio flux ratio. The lack of
change in the radio data now gives an upper limit on the clump mass.
The optical variability suggests that this may be an example of
stellar microlensing --- the clump(s) may simply be one or more
stars in the lens galaxy. I restrict attention to models where a
single clump is responsible for the variability, motivated by two
reasons.  First, the main goal is to determine whether sub-lensing
can explain image B, and it makes sense to start with the simplest
possible model to see whether it is sufficient. Second, the image
lies far from the center of the lens galaxy where the optical depth
for microlensing is small, which is equivalent to saying that
sub-lensing is likely to involve a single star.

\reffig{Bchi} shows the $\chi^2$ versus the clump mass for models
optimized over the position and velocity of the clump. The data can
be fit by a single point mass with any mass
$M \lesssim 200\ \hMsun$. (With a reasonable optical source size of
$10^{-4}$ mas [see \S 3], finite source size effects do not
preclude point mass clumps down to at least $0.1\ \hMsun$.) Thus,
the current data are consistent with microlensing by a single star
with a reasonable mass. The data can also be fit by an SIS clump
with $100 \lesssim M \lesssim 4000\ \hMsun$, although in this mass
regime the SIS model may be physically implausible.

\begin{figurehere}
\centerline{\epsfxsize=3.2in \epsfbox{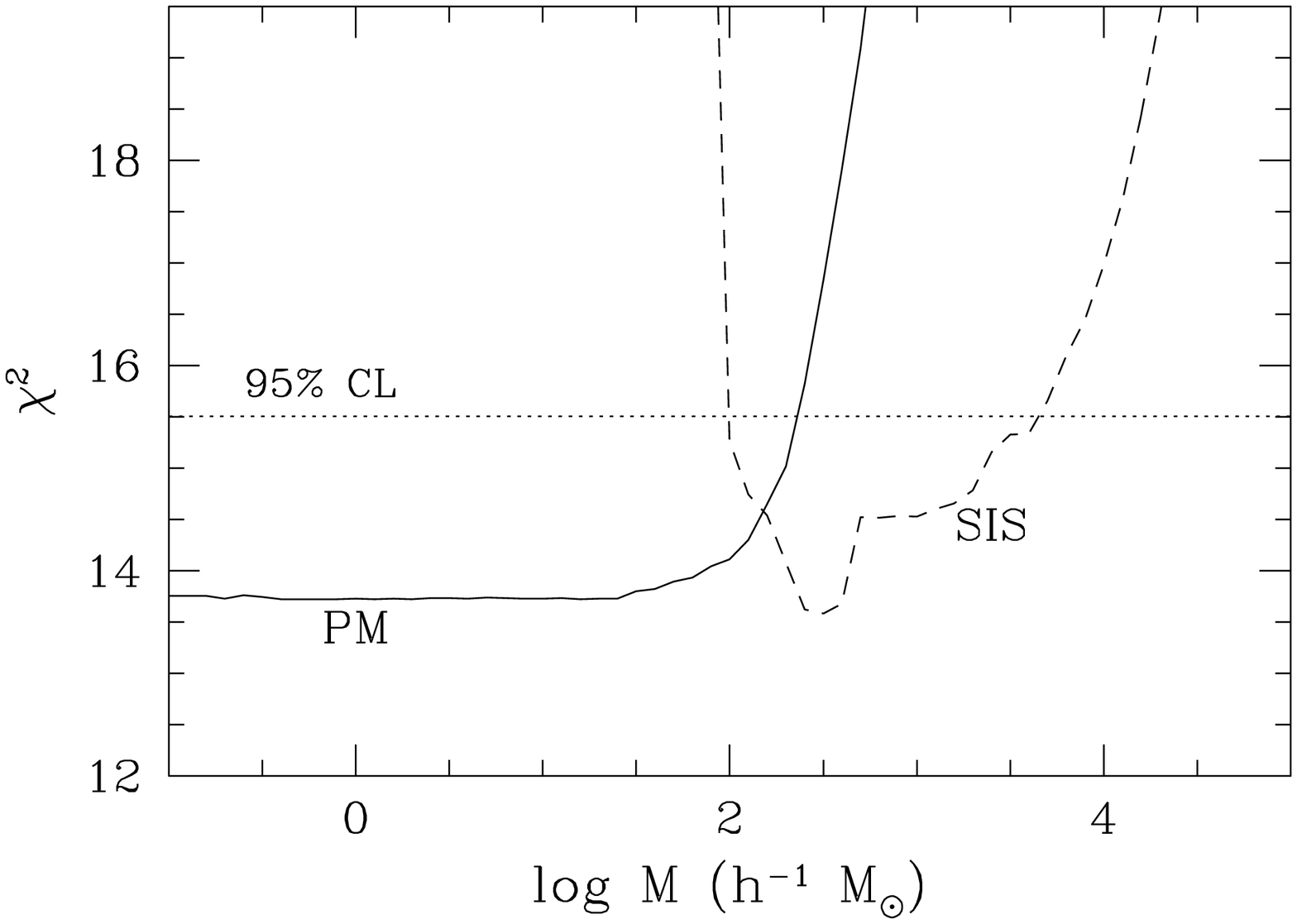}}
\caption{
The $\chi^2$ versus clump mass for sub-lensing models of image B.
The models are optimized over the position and velocity of the
clump, so there are 8 degrees of freedom. Results are shown for
point mass (PM) and SIS clump models. The dotted line shows the
95\% confidence limit.
}\label{fig:Bchi}
\end{figurehere}
\vspace{0.2cm}

\begin{figurehere}
\centerline{\epsfxsize=3.2in \epsfbox{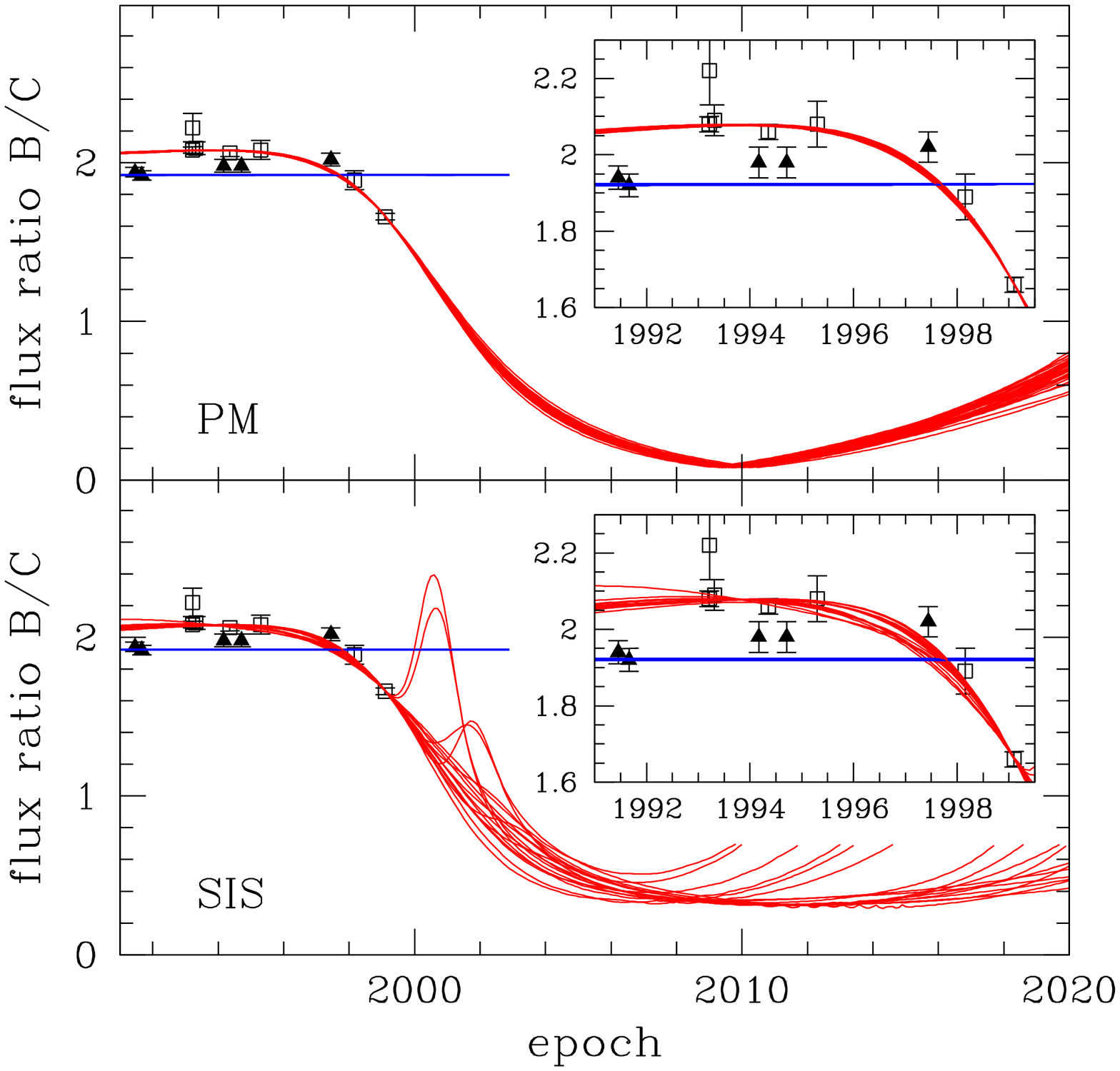}}
\caption{
The B/C flux ratio as a function of time. The points show the data,
while the curves show predictions from all of the models in
\reffig{Bchi} that fit the data at better than 95\% confidence. The
triangles and horizontal lines denote the radio flux ratio, while
the squares and curves denote the optical flux ratio. The main
panels show an extended time period, while the insets highlight the
period spanned by the data.
}\label{fig:lcrv2}
\end{figurehere}
\vspace{0.2cm}

\reffig{lcrv2} shows the B/C flux ratio as a function of time for
all of the models that fit the data at better than 95\% confidence.
Some SIS sub-lensing models predict that image B experienced a
caustic-crossing event in the year 2000 or 2001. More robustly,
all models predict that the optical flux ratio will continue to
decline for several more years, and will soon reach a point where
image B is fainter than both images A and C. The fact that
sub-lensing {\it de-amplifies\/} optical image B is important,
and is due to the fact that B is a parity-reversed image.  Further
monitoring of the B/C optical flux ratio will be very important
to test the sub-lensing hypothesis for image B.  Measurements
several times a year for several years should reveal continued
variability, providing improved data and allowing new models to
determine whether the variability requires one or many stars.
Polarimetric and/or spectral monitoring of the images could even
provide constraints on the relative sizes of the continuum
emission and the absorption regions in the source (e.g., Belle
\& Lewis 2000).

\section{Conclusions}

Substructure in the lens galaxy offers the first successful explanation
of the wavelength and time dependence in the flux ratios of the lens
B1422+231.  The difference between the optical and radio flux ratios
of images A and C implies a mass clump in front of image A.  A highly
concentrated clump must have a mass of $\sim 10^{4}$--$10^{5}\ \hMsun$,
while a more extended clump must have a mass of $\sim 10^{6}$--$10^{7}\
\hMsun$. This is the first evidence for a particular object of mass
$\sim 10^{4}$--$10^{7}\ \hMsun$ lying in a distant galaxy and detected
by its mass. Sub-mas resolution radio maps should either confirm the
clump and strongly constrain its mass, position, size, and density ---
or else rule out the clump hypothesis altogether.

The time dependence in the optical flux ratio of images B and C
implies that a small mass clump is passing in front of image B; the
object could be a normal star, or an extended object of mass
$\sim 10^{3}\ \hMsun$.  Models with a single clump predict that the
B/C optical flux ratio will continue to decline for several more years,
so photometric monitoring will test the sub-lensing hypothesis and
reveal whether the variability is due to one star or many.

This analysis of B1422+231 complements the recent statistical analyses
of lensing and substructure by Dalal \& Kochanek (2001) and Chiba
(2001).  The statistical approach uses flux ratios for an ensemble of
lenses to place limits on the statistical properties of subhalo
populations, showing that they are consistent with CDM and
inconsistent with known satellite populations.  By contrast,
B1422+231 demonstrates that analysis of more detailed data in
individual lenses, including photometry at multiple wavelengths and
epochs and high-resolution maps of radio images, can constrain
individual mass clumps in distant galaxies.

\acknowledgements
Support for this work was provided by NASA through Hubble
Fellowship grant HST-HF-01141.01-A from the Space Telescope Science
Institute, which is operated by the Association of Universities for
Research in Astronomy, Inc., under NASA contract NAS5-26555.


\begin{references}

\reference{}
Belle, K. E., \& Lewis, G. F. 2000, \pasp, 112, 320

\reference{}
Bode, P., Ostriker, J. P., \& Turok, N. 2001, \apj, 556, 93

\reference{}
Bullock, J. S., Kravtsov, A. V., \& Weinberg, D. H. 2000, \apj, 539, 517

\reference{}
Chang, K., \& Refsdal, S. 1979, \nat, 282, 561

\reference{}
Chiba, M. 2001, preprint astro-ph/0109499

\reference{}
Colin, P., Avila-Reese, V., \& Valenzuela, O. 2000, \apj, 542, 622

\reference{}
Dalal, N., \& Kochanek, C. S. 2001, preprint astro-ph/0111456

\reference{}
Falco, E. E., Impey, C. D., Kochanek, C. S., Leh\'ar, J., McLeod, B. A.,
Rix, H.-W., Keeton, C. R., Mu\~noz, J. A., \& Peng, C. Y. 1999, \apj, 523, 617

\reference{}
Hammer, F., Rigaut, F., Angonin-Willaime, M.-C., \& Vanderriest, C.
1995, \aap, 298, 737

\reference{}
Hogg, D. W., \& Blandford, R. D. 1994, \mnras, 268, 889

\reference{}
Impey, C. D., Foltz, C. B., Petry, C. E., Browne, I. W. A., \& Patnaik, A. R.
1996, \apj, 462, L53

\reference{}
Irwin, M. J., Webster, R. L., Hewett, P. C., Corrigan, R. T., \&
Jedrzejewski, R. I. 1989, \aj, 98, 1989

\reference{}
Keeton, C. R., Kochanek, C. S., \& Seljak, U. 1997, \apj, 482, 604

\reference{}
Klypin, A., Kravtsov, A. V., Valenzuela, O., \& Prada, F. 1999, \apj, 522, 82

\reference{}
Kundi\'c, T., Hogg, D. W., Blandford, R. D., Cohen, J. G., Lubin, L. M., \&
Larkin, J. E. 1997, \aj, 114, 2276

\reference{}
Mao, S., \& Schneider, P. 1998, \mnras, 295, 587

\reference{}
Metcalf, R. B., \& Madau, P. 2001, preprint astro-ph/0108224

\reference{}
Metcalf, R. B., \& Zhao, H. 2001, preprint astro-ph/0111427

\reference{}
Moore, B., Ghigna, S., Governato, F., Lake, G., Quinn, T., Stadel, J.,
\& Tozzi, P. 1999, \apj, 524, L19

\reference{}
Patnaik, A. R., Browne, I. W. A., Walsh, D., Chaffee, F. H., \& Foltz, C. B.
1992, \mnras, 259, 1P

\reference{}
Patnaik, A. R., Kemball, A. J., Porcas, R. W., \& Garrett, M. A. 1999,
\mnras, 307, L1

\reference{}
Patnaik, A. R., \& Narasimha, D. 2001, preprint astro-ph/0106104

\reference{}
Rees, M. J. 1984, \araa, 22, 471

\reference{}
Remy, M., Surdej, J., Smette, A., \& Claeskens, J.-F. 1993, \aap, 278, L19

\reference{}
Schneider, P., Ehlers, J., \& Falco, E. E. 1992, Gravitational Lenses
(New York: Springer)

\reference{}
Spergel, D. N., \& Steinhardt, P. J. 2000, Phys Rev Lett, 84, 3760

\reference{}
Wyithe, J. S. B., Webster, R. L., Turner, E. L., \& Mortlock, D. J.
2000, \mnras, 315, 62

\reference{}
Yee, H. K. C., \& Bechtold, J. 1996, \aj, 111, 1007

\end{references}
\end{document}